\documentclass[11pt,a4paper]{article}
\usepackage{amsmath,amssymb}
\usepackage{epsfig,graphicx}

\begin{document}

\title{\bf Perihelion precession and deflection of light in\\
       gravitational field of wormholes}
\author{V.~Strokov$^a$\footnote{{\bf e-mail}: strokov@asc.rssi.ru},
S.~Repin$^b$\footnote{{\bf e-mail}: repin@mx.iki.rssi.ru}
\\
$^a$ \small{\em Astro Space Center of P.~N.~Lebedev Physical Institute, Russian Academy of Sciences} \\
\small{\em ul. Profsoyuznaya 84/32, 117997 Moscow, Russian
Federation}
\\
$^b$ \small{\em Space Research Institute, Russian Academy of Sciences} \\
\small{\em ul. Profsoyuznaya 84/32, 117997 Moscow, Russian
Federation}}
\date{}
\maketitle

\newcommand{\Arcosh}{\mathop{\rm Arcosh}\nolimits}

\begin{abstract}
Quite exotic relativistic objects known as wormholes are
hypothetical candidates for central machine of active galactic
nuclei as well as black holes. We find the magnitude of the
perihelion precession and the deflection of light in gravitational
field of a wormhole and compare them with those for a black hole.
The impact parameter is taken to be much larger than the wormhole
throat size. We show that the relative difference between results
for a black hole and a wormhole may be significant and amount to
tens of percent.
\end{abstract}

\section{Introduction}
Einstein and Rosen \cite{einstein} were first to propose
nontrivial topological configurations known as bridges in the
frame\-work of the General Relativity (GR). Nowadays, nontrivial
configurations known as wormholes are widely discussed. Moreover,
the interest to these objects increased last decade.

There are several types of wormholes depending on their
topology~\cite{visser}. In this paper we concentrate on so-called
traversable Lorentzian wormholes~\cite{morris_1988}. Bodies can
freely pass through a traversable wormhole throat in both
directions and there are no event horizons in these objects. It
have been recently proposed~\cite{kardashev_2006} that observed
active galactic nuclei (AGN) and some other high-energy objects in
the Universe may be the former or present entrances to wormholes.
In light of this suggestion it is interesting to calculate some
effects which could possibly distinguish a wormhole from a black
hole in an experiment. Some of the effects, such as magnetic field
of a wormhole~\cite{kardashev_2006}, transition of light through a
wormhole throat~\cite{shatskiy}, lensing by
wormholes~\cite{lenses}, accretion onto wormholes
\cite{gonzalez_06} etc. have been proposed.

In this paper we calculate two classical GR effects, the
precession of perihelion (we still use ''perihelion'' for the
closest orbit point) and deflection of light in the gravitational
field of a wormhole with central symmetry, and compare them with
those for a black hole. Deflection of light was also calculated in
the paper \cite{shatskiy}, but with no relation to experiment. For
our calculations we use an explicit metrics of a wormhole made of
certain matter~\cite{kardashev_2006}. We also make a remark on the
third GR effect, gra\-vi\-ta\-tional redshift.

The outline of the paper is as follows. In Sect.~\ref{equations of
motion} we present the equations of motion in a spherically
symmetric wormhole field. In Sect.~\ref{perihelion} we find the
perihelion precession magnitude in two different ways. In
Sect.~\ref{deflection} we find the light deflection magnitude. And
in Sect.~\ref{discussion} we discuss the results and draw the
conclusions comparing the results with those for a black hole.

\section{\label{equations of motion}Equations of motion}

A quite general expression for the wormhole metrics with central
symmetry is \cite{morris_1988}:

\begin{equation}
\label{general metrics}
   ds^2 = e^{2\phi(r)} dt^2 -
          \frac{dr^2}{\displaystyle 1 - \frac{b(r)}{r}} -
           r^2 d\Omega^{2},
\end{equation}
where $d\Omega^{2}=\sin^{2}\theta d\varphi^{2}+d\theta^{2}$,
$b(r)$ and $\phi(r)$ are the functions of a radial coordinate
only.

In order to construct a wormhole a phantom matter
($p<-\varepsilon$) with anisotropic equation of state is
required~\cite{lobo_05}. We consider the following equation of
state~\cite{kardashev_2006}:
\begin{equation}
1+\delta=-p_{\parallel}/\varepsilon=p_{\perp}/\varepsilon,
\end{equation}
where $\delta>0$. The possibility of realization of this equation
of state is not clear yet, nevertheless it is widely considered.
Under these assumptions on what the wormhole is made of, in the
paper \cite{kardashev_2006} the metrics was found explicitly:
\begin{eqnarray}
\label{the metrics}
ds^{2}&=& \left(1-\frac{r_{h}}{r}\right)^{2+2\delta}c^{2}dt^{2}-\nonumber\\
&&{}-\frac{dr^{2}}{1-\displaystyle\frac{r_{h}}{r}\left[1+
\left(1-\frac{r_{h}}{r}\right)^{1-\delta}\right]}-r^{2}d\Omega^{2}.
\end{eqnarray}
In the limit case of $\delta=0$, the metrics turns into the
Reissner-Nordstr\"{o}m (RN) metrics, and $r_h= GM/c^2$ is a
gravitational radius of a corresponding RN black hole.

  We use the metrics (\ref{the metrics}) to derive the equations
of motion and then to carry out calculations of the perihelion
precession and the deflection of light. The metrics (\ref{the
metrics}) possesses spherical symmetry, hence, we can consider
only the equatorial plane motion and set $\theta=\pi/2$. Thus,
$d\Omega^{2}=d\varphi^{2}$.

We can write the Hamilton-Jacobi equation for a particle with the
mass $m$ as following:
\begin{equation}
g^{ik}\frac{\partial S}{\partial x^i}
      \frac{\partial S}{\partial x^k} = m^{2}c^{2}  \qquad\qquad
      (i,k = 0,1,2,3).
\end{equation}
For the metrics (\ref{the metrics}) the latter equation takes the
form:
\begin{equation}
\label{H-J}
\begin{array}{l}
  \displaystyle\left(1-\frac{r_{h}}{r}\right)^{-2-2\delta}
  \left(\frac{\partial S}{c\partial t}\right)^{2}-
  \frac{1}{r^{2}}
    \left(\frac{\partial S}{\partial \varphi}\right)^{2}- \\
    \\
    -\displaystyle\left(1-\frac{r_{h}}{r}\right)
  \left[1-\frac{r_{h}}{r}
    \left(1-\frac{r_{h}}{r}\right)^{-\delta}
  \right]
  \left(\frac{\partial S}{\partial r}\right)^{2} =
  m^{2}c^{2}.
\end{array}
\end{equation}
Since the metrics does not depend explicitly on time $t$ and the
angle $\varphi$, we look for a solution of equation (\ref{H-J}) in
the form:
\begin{equation}
\label{sol-form}
   S = -Et + L\varphi + S_r(r),
\end{equation}
where $E$ and $L$ are conserving energy and projection of angular
momentum on $z$ axis, respectively. Substituting
form~(\ref{sol-form}) to equation~(\ref{H-J}) we find $S_r(r)$ and
obtain the solution:
\begin{equation}
\label{action}
\begin{array}{l}
S=-Et+L\varphi{}+\\
\\
+{}\displaystyle\int\sqrt{\frac{\displaystyle\left(1-
\frac{r_{h}}{r}\right)^{-2-2\delta}\frac{E^{2}}{c^{2}}-\frac{L^{2}}{r^{2}}-
m^{2}c^{2}}{\displaystyle\left(1-\frac{r_{h}}{r}\right)\left[1-
\frac{r_{h}}{r}\left(1-\frac{r_{h}}{r}\right)^{-\delta}\right]}}\,dr.
\end{array}
\end{equation}

   Substituting then the partial derivatives into the equalities
\begin{equation}
    p^i = m \frac{dx^i}{ds} = g^{ik}p_k =
       - g^{ik} \frac{\partial S}{\partial x^k}
       \label{four_momentum}
\end{equation}
we find the equations of motion explicitly:
\begin{eqnarray*}
  m \frac{dt}{ds}  & = &
     \frac{E}{c^2\left(1 - \displaystyle\frac{r_h}{r}\right)^{2+2\delta}}, \\
  m \frac{dr}{ds}  & = &
     \left[\frac{\displaystyle\mathstrut E^2}
        {\displaystyle c^2\left(1 - \frac{r_h}{r}\right)^{2+2\delta}} -
             \frac{\displaystyle L^2}{\displaystyle r^2} -
             m^2c^2
     \right]^{1/2}\times \\
     &&\times\left[
        1 - \frac{r_h}{r}
          \left(
            1 +
            \left( 1 - \frac{r_h}{r} \right)^{1 - \delta}
          \right)
     \right]^{1/2}, \\
  m \frac{d\varphi}{ds}  & = & \frac{L}{r^2}.
\end{eqnarray*}

  Note that we consider the motion in the equatorial plane so
that the fourth equation (for $\theta$) is unnecessary.

\section{\label{perihelion}Perihelion precession}
\subsection{Trajectory based consideration}
\label{trajectory}

We suggest that one revolution is a motion of a body from pericenter
to pericenter. If the orbit is not closed the change of angle
$\varphi$, which corresponds to this motion may be both grater or
less than $2\pi$. As usual \cite{mechanics}, the orbit equation
comes from the relation: \mbox{$\partial S/\partial
L=const=\varphi_{0}$}. Thus, the following formula yields the
perihelion precession $\Delta\varphi$:
\begin{equation}
\label{delta-phi}
\pi+\frac{\Delta\varphi}{2}=\int_{r_{min}}^{r_{max}}\frac{Ldr}{r^{2}\sqrt{A(r)}},
\end{equation}
where
$$
\begin{array}{l}
A(r)=\displaystyle \frac{\displaystyle 1-\frac{r_{h}}{r}\left(1-
\frac{r_{h}}{r}\right)^{-\delta}}{\displaystyle
\left(1-\frac{r_{h}}{r}\right)^{1+2\delta}}\frac{E^{2}}{c^{2}} \,-\\
\\
-\displaystyle\left(\frac{L^{2}}{r^{2}}+m^{2}c^{2}\right)\left(1-\frac{r_{h}}{r}\right)\left[1-
\frac{r_{h}}{r}\left(1-\frac{r_{h}}{r}\right)^{-\delta}\right].
\end{array}
$$
The energy $E$ also includes the rest mass. We make the
substitution $E\rightarrow E+mc^{2}$ to exclude the rest mass from
$E$ and obtain:
\begin{eqnarray}
\label{under-root} A(r)&=& \displaystyle\frac{\displaystyle
1-\frac{r_{h}}{r}\left(1-
\frac{r_{h}}{r}\right)^{-\delta}}{\displaystyle
\left(1-\frac{r_{h}}{r}\right)^{1+2\delta}}\left(\frac{E^{2}}{c^{2}}+2Em\right)-\nonumber\\
&-& \frac{L^{2}}{r^{2}}\left(1-\frac{r_{h}}{r}\right)\left[1-
\frac{r_{h}}{r}\left(1-\frac{r_{h}}{r}\right)^{-\delta}\right]+\nonumber\\
&+&\displaystyle m^{2}c^{2}\left[\frac{1}{\displaystyle\left(1-
\frac{r_{h}}{r}\right)^{1+2\delta}}-1+\frac{r_{h}}{r}\right]\times\nonumber\\
&\times &\left[1-
\frac{r_{h}}{r}\left(1-\frac{r_{h}}{r}\right)^{-\delta}\right].
\end{eqnarray}

Since we consider the long-distance motion, i.e. $r_{h}/r\ll 1$,
we should retain only the terms that are not less than
$(r_{h}/r)^{2}$. Taking into account the estimations:
\begin{equation}
\label{estimation}
\frac{E}{mc^{2}}\sim\frac{L^{2}}{m^{2}c^{2}r^{2}}\sim\frac{r_{h}}{r},
\end{equation}
we obtain with required accuracy:
\begin{eqnarray*}
A(r)&=&
2m\left(E+\frac{(mc^{2}(1+\delta)+2E\delta)r_{h}}{r}\right)-\frac{L^{2}}{r^{2}}+\\
&+&\frac{E^{2}}{c^{2}}+\frac{(1+\delta)(2\delta-1)m^{2}c^{2}r_{h}^{2}}{r^{2}}+\frac{2L^{2}r_{h}}{r^{3}},
\end{eqnarray*}
or
\begin{equation}
A(r)=2m\left(E+\frac{\alpha}{r}\right)-\frac{L^{2}}{r^{2}}+
\frac{E^{2}}{c^{2}}-\frac{2m\beta}{r^{2}}-\frac{2m\gamma}{r^{3}},
\end{equation}
where
\begin{eqnarray}
\label{coeffs} \alpha &=&(mc^{2}(1+\delta)+2E\delta)r_{h}, \nonumber\\
\beta
&=&-\displaystyle\frac{1}{2}(1+\delta)(2\delta-1)mc^{2}r_{h}^{2},
\nonumber\\ \gamma &=&-\displaystyle\frac{L^{2}r_{h}}{m}.
\end{eqnarray}

In this section we calculate $\Delta\varphi$ by expanding the
integral in equation (\ref{delta-phi}) with respect to the small
terms $\beta/r^{2}$ and $\gamma/r^{3}$~\cite{mechanics}. Thus, we
obtain
\begin{equation}
\Delta\varphi=-\frac{2\pi\beta m}{L^{2}}-\frac{6\pi\alpha\gamma
m^{2}}{L^{4}}.
\end{equation}
Using relations~(\ref{coeffs}) we have:
\begin{eqnarray*}
\Delta\varphi &=&\frac{(1+\delta)(2\delta+5)\pi
m^{2}c^{2}r_{h}^{2}}{L^{2}}\times\\
&\times&
\left(1+\frac{12\delta}{(1+\delta)(2\delta+5)}\frac{E}{mc^{2}}\right).
\end{eqnarray*}
The second term in the brackets is small in the framework of the
considered approximation and should be omitted. Then we have
\begin{equation}
\Delta\varphi = \frac{(1+\delta)(2\delta+5)\pi
                 m^{2}c^{2}r_{h}^{2}}{L^{2}}.
   \label{Delta_phi}
\end{equation}

\subsection{Action based consideration}

To calculate the perihelion precession we can also follow the
method developed in \cite{fieldtheory} and consider the action $S$
instead of the equations of motion or the trajectory.

Let us denote $x=r_{h}/r$. The factor in the term with $L^2/r^2$
under the square root in equation~(\ref{action}) is
\begin{equation}
  f_1(x) = \left(1 - x\right)^{-1}
             \left[1 - x\right(1 - x\left)^{-\delta}
             \right]^{-1}
        \label{f_1_function}
\end{equation}
Expansion to the linear order yields
\begin{equation}
  f_1(x) = 1 + 2x + \dots
        \label{f_1_series}
\end{equation}
Then the term with $L^2/r^2$ in the integral in equation
(\ref{action}) is approximately
\begin{equation}
  -\frac{L^2}{r^2} \cdot \left(1 + 2\frac{r_h}{r}\right).
\end{equation}
Then we introduce a new variable $r^\prime$:
\begin{equation}
\label{new-var}
  \frac{1}{r^{\prime 2}} = \frac{1}{r^2} \cdot \left(1 +
  2\frac{r_h}{r}\right),
\end{equation}
or
\begin{equation}
\label{new-var2}
                  r = r^\prime + r_h.
\end{equation}
In terms of $r^\prime$ the term with $L^2/r^2$ under the integral
takes the form  $- \,L^2/r^{\prime 2}$. Then we calculate how the
change of variables (\ref{new-var2}) influences other terms. One
should first expand each factor in equation (\ref{action}) to the
second order:
\begin{equation}
\begin{array}{l}
   \displaystyle\left(1 - \frac{r_h}{r^\prime + r_h}\right)^{-1}
     \left[ \,
       1 - \frac{r_h}{r^\prime + r_h}
       \left(1 -
          \frac{r_h}{r^\prime + r_h}
       \right)^{-\delta}
     \right]^{-1}
     \sim \\
     \\
     \sim\displaystyle
     1 + 2\,\frac{r_h}{r^\prime} +
     \left(1 + \delta\right) \cdot \frac{r_h^2}{r^{\prime 2}}
\end{array}
\end{equation}
\begin{equation}
\begin{array}{l}
   \displaystyle\left(1 - \frac{r_h}{r^\prime + r_h}\right)^{-3-2\delta}
     \left[ \,
       1 - \frac{r_h}{r^\prime + r_h}
       \left(1 -
          \frac{r_h}{r^\prime + r_h}
       \right)^{-\delta}
     \right]^{-1}
     \sim
     \\
     \\
     \sim\displaystyle
     1 + \left(4 + 2\delta\right)\cdot\frac{r_h}{r^\prime} +
     \left(2\delta^2 + 8\delta + 6\right) \cdot
      \frac{r_h^2}{r^{\prime 2}}
\end{array}
\end{equation}

As in the trajectory method (subsection \ref{trajectory}) we
exclude the rest mass by substitution  $E \to E + mc^2$ and for
the sake of brevity omit the prime in $r^\prime$. Then the
under-root expression takes the form:
\begin{equation}
\begin{array}{l}
\displaystyle\frac{(E + mc^2)^2}{c^2}
 \left[
  1 + \left(4 + 2\delta\right)\cdot\frac{r_h}{r} +
     \left(2\delta^2 + 8\delta + 6\right) \cdot
      \frac{r_h^2}{r^2}
 \right] -\\
 \\
 -\displaystyle\frac{L^2}{r^2} -
 m^2c^2
 \left[
  1 + 2\,\frac{r_h}{r} +
     \left(1 + \delta\right) \cdot \frac{r_h^2}{r^2}
 \right].
\end{array}
\end{equation}

After opening the brackets, ordering the degrees of $r_h/r$ and
taking into account estimations~(\ref{estimation}) and the fact
that $E \ll mc^2$ and $r_h/r \ll 1$, the latter expression can be
written with required accuracy~as:
$$
\begin{array}{l}
 \displaystyle\left(\frac{E^2}{c^2} + 2Em\right)+2
 \left[
  2Em\left( 2+\delta \right) + m^2c^2\left( 1+\delta \right)
 \right]
 \frac{r_h}{r} \, +\\
 \\
 + \, \displaystyle m^2c^2\left( 1+\delta \right)\left( 5+2\delta \right)
 \frac{r_h^2}{r^2} -
 \frac{L^2}{r^2}.
\end{array}
$$
The action $S$ itself takes the form:
\begin{eqnarray}
\label{expanded action}
 S &=& -Et + L\varphi + \int
  \left[
    \left(\frac{E^2}{c^2} + 2Em\right) \right.
 + \nonumber\\
    &+& 2
    \left[
    2Em\left( 2+\delta \right) + m^2c^2\left( 1+\delta \right)
  \right]
  \frac{r_h}{r}-\nonumber\\
    &-& \left. \frac{1}{r^2}
    \left(
      L^2 -
      m^2c^2\left( 1+\delta \right)\left( 5+2\delta \right) r_h^2
    \right)
  \right]^{1/2}dr.
\end{eqnarray}

As it is well known, the correction factors in first two terms in
the integrand in equation (\ref{expanded action}) cause only the
correspondence between the particle energy, angular momentum and
parameters of its Kepler ellipse. The change in factor in front of
$1/r^2$ leads to the systematic {\it secular} precession of the
orbit perihelion.

As in the first method, the trajectory is determined by the
equation:
\begin{equation}
   \varphi + \frac{\partial S_r}{\partial L} = const,
\end{equation}
where $S_r(r)$ is a radial part of the action (i.e. the integral
in eq. (\ref{expanded action})). Hence, the change of angle
$\varphi$ after one revolution (from perihelion to perihelion) is
\begin{equation}
  2\pi + \Delta\varphi = - \frac{\partial}{\partial L} \Delta S_r,
\end{equation}
where $\Delta S_r$ is the respective change of $S_r$. Expanding
$S_r$ with respect to the small correction factor in front of
$1/r^2$, we obtain:
\begin{equation}
  \Delta S_r = \Delta S_r^{(0)} -
   \frac{m^2c^2\left( 1+\delta \right)\left( 5+2\delta \right) r_h^2}{2L}
   \cdot
   \frac{\partial \Delta S_r^{(0)}}{\partial L}.
\end{equation}
Then we differentiate the latter relation with respect to $L$ and
take into account that
\begin{equation}
  -\, \frac{\partial}{\partial L} \Delta S_r^{(0)} =
      \Delta\varphi^{(0)} = 2\pi.
\end{equation}
We also neglect the second derivative $\partial^2\Delta
S_r^{(0)}/\partial L^2$. As a result, we find:
\begin{equation}
 \Delta\varphi =
    \frac{
      \left( 1+\delta \right)
      \left( 5+2\delta \right)\pi m^2c^2 r_h^2}{L^2}.
\end{equation}
Certainly, this result coincides with
expression~(\ref{Delta_phi}).

\section{\label{deflection}Deflection of light}
To calculate the deflection of light we also use the
Hamilton-Jacobi method. Henceforth, we follow the technique set
out in \cite{fieldtheory}. We first write the Hamilton-Jacobi
equation for the eikonal $\psi$ (obviously, $m=0$ for light):
\begin{equation}
g^{ik}\frac{\partial\psi}{\partial
x^i}\frac{\partial\psi}{\partial x^k}=0.
\end{equation}
As in Sect. 2 we look for a solution in the form:
\begin{equation}
\label{sol-form1} \psi=-\omega_{0}t+L\varphi+\psi_{r}(r),
\end{equation}
where $\omega_{0}$ is frequency of the light observed at infinity.

Thus, for $\psi_{r}$ we obtain:
\begin{equation}
\psi_{r}=\frac{\omega_{0}}{c}\int\sqrt{\frac{\displaystyle\left(1-
\frac{r_{h}}{r}\right)^{-2-2\delta}-\frac{\rho^{2}}{r^{2}}}
{\displaystyle\left(1-\frac{r_{h}}{r}\right)\left[1-
\frac{r_{h}}{r}\left(1-\frac{r_{h}}{r}\right)^{-\delta}\right]}}dr,
\end{equation}
where the notation $\rho=cL/\omega_{0}$ is introduced. The
expansion of the integrand to the first order in $r_{h}/r$ yields
\begin{equation}
\psi_{r}=\frac{\omega_{0}}{c}\int\sqrt{1-
\frac{\rho^{2}}{r^{2}}+2\frac{r_{h}}{r}\left((2+\delta)-\frac{\rho^{2}}{r^{2}}\right)}dr.
\end{equation}
Expanding the integral with respect to $r_h/r$ we further obtain:
\begin{eqnarray*}
\psi_{r}&=&\psi_{r}^{(0)}+
\frac{(2+\delta)r_{h}\omega_{0}}{c}\int\frac{dr}{\sqrt{r^{2}-\rho^{2}}}-\\
&-&
\frac{r_{h}\rho^{2}\omega_{0}}{c}\int\frac{dr}{r^{2}\sqrt{r^{2}-\rho^{2}}},
\end{eqnarray*}
where $\psi_{r}^{(0)}$ corresponds to the free (straight)
propagation of the light. Evaluating the integrals with the limits
from $\rho$ to a large distance $R$, we obtain the eikonal change
$\Delta\psi_{r}$ explicitly:
\begin{eqnarray}
\label{eikonal} \Delta\psi_{r}&=& \Delta\psi_{r}^{(0)}+
\frac{2(2+\delta)r_{h}\omega_{0}}{c}\Arcosh{\frac{R}{\rho}}-\nonumber\\
&-& \frac{2r_{h}\omega_{0}}{c}\sqrt{1-\frac{\rho^{2}}{R^{2}}}.
\end{eqnarray}
From equation (\ref{sol-form1}) one can see that the light
deflection $\Delta\theta$ is
\begin{equation}
\Delta\theta=-\frac{\partial\Delta\psi_{r}}{\partial L}.
\end{equation}
Using equation (\ref{eikonal}) and making $R$ infinite after
evaluation the derivative with respect to $L$ we finally obtain:
\begin{equation}
\label{light-deflection}
\Delta\theta=\frac{2(2+\delta)r_{h}}{\rho}.
\end{equation}

\section{\label{discussion}Discussion}

According to the paper~\cite{kardashev_2006} $r_h=GM/c^2$. In the
classical limit the metrics (\ref{the metrics}) yields Newtonian
potential $\alpha/r$, where $\alpha=(1+\delta)GMm$ (see
expr.~(\ref{coeffs})). This implies that an observer would measure
that the body orbits a gravitating center with the mass
\mbox{$M_{0}=(1+\delta)M$}. By the way, this potential $\alpha/r$
immediately yields the gravitational redshift, viz. $\Delta\nu/\nu
= -GM_0/c^2r_0$, for a photon emitted at radius $r_0$ and
registered at infinity. However, this effect does not allow us to
distinguish a wormhole from a black hole.

Finally,
\begin{equation}
\label{final formula} \Delta\varphi=
\frac{2\delta+5}{1+\delta}\frac{\pi
G^{2}M_{0}^{2}m^{2}}{c^{2}L^{2}},
\end{equation}
where $M_0$ is a mass of a wormhole, measured by a distant
observer.

The perihelion precession $\Delta\varphi_{BH}$ for a body orbiting
a black hole with mass $M_{0}$ is given by the
formula~\cite{fieldtheory}:
\begin{equation}
\Delta\varphi_{BH}=\frac{6\pi G^{2}M_{0}^{2}m^{2}}{c^{2}L^{2}}.
\end{equation}
Making use of equation~(\ref{final formula}) we obtain the ratio
\begin{equation}
\frac{\Delta\varphi}{\Delta\varphi_{BH}}=\frac{2\delta+5}{6(1+\delta)}=
\frac{1}{3}+\frac{1}{2(1+\delta)}.
\end{equation}
Usually $\delta$ is considered to be small and
positive~\cite{kardashev_2006}. We assume $0<\delta<1$. Then, from
the latter equation it is clear that $\Delta\varphi$ is always
smaller than $\Delta\varphi_{BH}$. $\Delta\varphi$ ranges in the
interval:
\begin{equation}
0.6\Delta\varphi_{BH}\lesssim\Delta\varphi\lesssim
0.8\Delta\varphi_{BH}.
\end{equation}
It means that the smallest difference is about $20$\%.

Analogously, using the experimentally measured mass $M_{0}$ in the
light deflection formula~(\ref{light-deflection}), we obtain:
\begin{equation}
\label{final formula 1}
\Delta\theta=\frac{2+\delta}{1+\delta}\frac{r_{g}}{\rho},
\end{equation}
where $r_{g}=2GM_{0}/c^{2}$ (as if the gravitating center were a
BH with gravitational radius $r_g$). The black hole light
deflection amounts to~\cite{einstein-1, fieldtheory}:
\begin{equation}
\Delta\theta_{BH}=2\frac{r_{g}}{\rho}.
\end{equation}
Hence, we obtain for the worm to black hole ratio:
\begin{equation}
\frac{\Delta\theta}{\Delta\theta_{BH}}=\frac{2+\delta}{2(1+\delta)}=
\frac{1}{2}+\frac{1}{2(1+\delta)}.
\end{equation}
Therefore, $\Delta\theta$ ranges in the interval
\begin{equation}
0.75\Delta\theta_{BH}\leq\Delta\theta\leq\Delta\theta_{BH}.
\end{equation}
The difference may amount to $25$\%.

The magnitude of the calculated effects is of order of several
tens of percents. This gives a hope that in the not-so-distant
future these differences will possibly be registered, and one will
be able to answer the question whether the wormholes are
encountered or not in some astrophysical objects.

The deflection of light in the gravitation field of the Sun during
total eclipses was measured with at least one percent precision as
well as Mercury's perihelion precession. It means that most
probably there is no wormhole inside the Sun.

\section{Acknowledgements}
One of us (VS) acknowledges support from the Scientific Study
Complex of the P.N.~Lebedev Physics Institute and the Russian
Foundation for Basic Research (project code 07-02-00886). Another
author (SR) is very grateful to Prof.~E. V. Sta\-ro\-sten\-ko,
Dr.~R.~E.~ Be\-res\-ne\-va and Dr.~O.~N.~Su\-men\-ko\-va for the
possibility of fruitful working under this problem. We are thankful
to Prof.~V.N.~Lukash for discussion and useful comments.

\end{document}